\documentclass[12pt,a4paper]{article}
\usepackage[text={16.5cm,23cm},centering]{geometry}
\usepackage{amsmath}
\usepackage{amsthm}
\usepackage[utf8]{inputenc}
\usepackage{tgtermes,tgheros,tgcursor}
\usepackage[varg]{newtxmath}
\usepackage{hyperref}
\hypersetup{%
 colorlinks=true,%
 linkcolor=blue,
 citecolor=blue,
}
\usepackage{xspace}
\usepackage{cite}
\usepackage{comment}

\numberwithin{equation}{section}

\newcommand{\pdv}[1]{\frac{\partial}{\partial #1}}
\newcommand{\del}{\partial}

\newcommand{\bulk}{\ensuremath{M}\xspace}
\newcommand{\brane}{\ensuremath{Q}\xspace}
\newcommand{\bdy}{\ensuremath{\Sigma}\xspace}
\newcommand{\action}{I}
\newcommand{\couple}{q}
\newcommand{\AdS}{\ensuremath{\text{AdS}_3}\xspace}
\newcommand{\mass}{\alpha}
\newcommand{\Rcal}{\mathcal{R}}

\newcommand{\Iren}{I_{\mathrm{ren}}}

\begin{document}
\begin{center}
  \begin{flushright}
    OU-HET 1328
  \end{flushright}
  \vspace{6ex}
  {\Large \bfseries \boldmath $\text{AdS}_3/\text{BCFT}_2$ correspondence and ’t Hooft anomaly of U(1) symmetry}\\
  \vspace{4ex}
  {\Large Hikaru Toya and Satoshi Yamaguchi}\\
  \vspace{2ex}
  {\itshape Department of Physics, Graduate School of Science, 
  \\
  The University of Osaka, Toyonaka, Osaka 560-0043, Japan}\\
  \vspace{1ex}
  \begin{abstract}
    We investigate how a U(1) 't Hooft anomaly obstructs symmetry-preserving boundary conditions from the perspective of the AdS\(_3\)/BCFT\(_2\) correspondence. By considering a U(1) Maxwell--Chern--Simons system in the bulk spacetime, we show that the BCFT current is not conserved at the boundary. We also derive the general solution for the gauge field in AdS\(_2\) slicing.
  \end{abstract}
\end{center}

\vspace{4ex}
\section{Introduction and summary}
Since its proposal, the AdS/CFT correspondence \cite{Maldacena:1997re} has been extensively studied in a wide variety of contexts. 
In particular, conformal field theories defined on manifolds with boundaries, known as boundary conformal field theories (BCFTs) \cite{Cardy:1984bb,Cardy:1989ir,Cardy:2004hm,McAvity:1993ue,McAvity:1995zd}, can be described holographically through the AdS/BCFT correspondence \cite{Takayanagi:2011zk,Fujita:2011fp}.
The AdS/BCFT correspondence is realized by extending the boundary of the BCFT into the AdS bulk.
The resulting boundary of the AdS spacetime is called the end-of-the-world (EoW) brane. 

Boundary conditions are one of the defining ingredients of BCFTs; for example, they are imposed on the energy-momentum tensor and on the currents associated with global symmetries.
From the perspective of the AdS/BCFT correspondence, the boundary conditions of the BCFT are encoded in the boundary conditions imposed on the EoW brane for the bulk fields that are dual to the BCFT operators. 

More generally, quantum field theories are constrained by their symmetries and the associated anomalies.
The choice of boundary conditions is also constrained by these symmetries and anomalies (e.g. \cite{Jensen:2017eof,Thorngren:2020yht,Choi:2023xjw}).
In the case of BCFTs, it is known that if a global symmetry of the BCFT possesses an anomaly, it is impossible to introduce a boundary while maintaining the conservation of the corresponding current at the boundary.

In holography, global symmetries and their anomalies in CFTs are encoded in the dynamics of bulk gauge fields.
For example, it has been known since the early days of the AdS/CFT correspondence that a bulk Chern--Simons term is dual to the anomaly of the boundary theory \cite{Witten:1998qj,Jensen:2010em}, providing a holographic realization of anomaly inflow \cite{Callan:1984sa}.
Furthermore, within the AdS/BCFT correspondence, it has been shown that the gauge symmetry in the bulk is closely related to the conservation of the current associated with the corresponding global symmetry of the BCFT at the boundary \cite{Suzuki:2024cqy}.
Gauge-field boundary conditions and Chern--Simons terms on the EoW brane have been also studied in the context of Hall transport in \(\text{AdS}_4/\text{BCFT}_3\) \cite{Fujita:2012fp}.

In this paper, we study the relation between boundaries and 't Hooft anomalies from the holographic perspective.
As mentioned above, both anomalies and boundaries admit holographic descriptions: anomalies in CFTs correspond to the Chern--Simons terms in the bulk, while the boundaries in CFTs correspond to EoW branes in the bulk.
It is therefore important to understand holographically how the Chern--Simons term leads to the breaking of the symmetry in the presence of a boundary.

Specifically, we consider a U(1) Maxwell--Chern--Simons field in the \(\text{AdS}_3/\text{BCFT}_2\) correspondence.
We impose Dirichlet boundary conditions on the EoW brane and an additional condition on the magnetic flux near the BCFT boundary.
We compute the BCFT current using an explicit family of solutions.
We find allowed solutions for which the normal component of the BCFT current is nonzero at the boundary.
Since current conservation requires this component to vanish for every allowed configuration, these solutions establish that the boundary conditions considered here do not preserve the U(1) symmetry.
This provides an explicit holographic realization of the obstruction associated with the 't Hooft anomaly \cite{Jensen:2017eof,Thorngren:2020yht,Choi:2023xjw}.

This family of solutions is sufficient to establish this violation of current conservation.
We also develop a mode decomposition to describe the broader solution space and identify how this family is embedded in it.

An important next step is to determine admissible boundary conditions for the gauge field on the EoW brane from explicit holographic models.
Such models would clarify how the boundary dynamics selects the bulk solutions and controls the current at the BCFT boundary.
Studies of Weyl-anomaly-induced chiral currents in \(\text{AdS}_5/\text{BCFT}_4\) provide examples of how gauge-field terms on the EoW brane determine boundary conditions \cite{Miao:2022oas,Zhao:2023rno}.
It would also be interesting to extend the analysis to non-Abelian gauge fields and examine how the corresponding anomalies constrain boundary charge transport.

The remainder of this paper is organized as follows.
In Section \ref{2}, we show, within the framework of the AdS/BCFT correspondence, that the BCFT current is not conserved at the boundary.
We consider the Maxwell--Chern--Simons theory in the bulk and explicitly derive solutions to its equations of motion.
We use these solutions to compute the BCFT current holographically and investigate its behavior at the boundary.
Finally, in Section \ref{3}, we study general solutions to the bulk equations of motion.

\section{A Family of Solutions and Current Non-conservation}
\label{2}
In this section, we explain the non-conservation of the current at the boundary of a BCFT with a ’t Hooft anomaly from the perspective of the AdS/BCFT correspondence.
Unlike in \cite{Suzuki:2024cqy}, we introduce an additional Chern--Simons term into the bulk action to account for the ’t Hooft anomaly. 
For the relationship between ’t Hooft anomalies and current conservation at the boundary from the viewpoint of quantum field theory, see \cite{Jensen:2017eof,Thorngren:2020yht,Choi:2023xjw}.
\subsection{Bulk geometry}
\label{2.1}
In this subsection, we describe the bulk background geometry of AdS/BCFT, mostly following \cite{Takayanagi:2011zk}.  We focus on the case in which the bulk spacetime is three-dimensional.

The action for the metric is given by
\begin{equation}
  I_{\text{background}} = I_{\text{bulk}} + I_{\text{brane}} + I_{\text{bdy}},
  \label{metricaction}
\end{equation}
where
\begin{align}
  I_{\text{bulk}}  &= \frac{1}{16 \pi G_N} \int_\bulk d^{3}x\sqrt{-g}\, (R - 2 \Lambda), \\
  I_{\text{brane}} &= \frac{1}{8 \pi G_N} \int_\brane d^{2}x\sqrt{-h}\, (K - T).
\end{align}
Here, $G_N$ is the Newton constant. \bulk denotes the three-dimensional bulk spacetime manifold. The boundary of \bulk is given by \(\bdy \cup \brane\),
where \bdy is the conformal boundary located at infinity, and \brane is the boundary at a finite distance, called ``the end of the world (EoW) brane''.
\(g\) and \(h\) denote the bulk metric and the induced metric on \brane, respectively.  $\Lambda<0$ is the cosmological constant.  Here we choose units in which $\Lambda =-1$ for notational simplicity.
$K$ is the trace of extrinsic curvature $K_{ab}$ for the EoW brane \brane. $K_{ab}$ is defined by $K_{\mu\nu}:=\nabla_{\mu}n_{\nu}$ projected to the boundary indices $a,b$, where $n_{\mu}$ is the outward-pointing unit normal vector of \brane.
$T$ is a constant, which is interpreted as the brane tension.
\(I_{\text{bdy}}\) denotes the standard boundary terms on the
asymptotic boundary \bdy, including the Gibbons--Hawking--York term \cite{Gibbons:1976ue,York:1972sj} and
local counterterms required by holographic renormalization \cite{Skenderis:2002wp}.  We will not
need their explicit form below.

When \brane is empty, the action \eqref{metricaction} has the \AdS solution. The metric is written in terms of the Poincaré coordinates $z,t,x$ as
\begin{align}
  ds^2 = \frac{dz^2-dt^2+dx^2}{z^2}.
\end{align}
The AdS radius is unity in these units.
Another set of coordinates $\rho, y,t$ is useful for describing AdS/BCFT. The two coordinate systems are related by
\begin{align}
  z=\frac{y}{\cosh \rho},\quad x= y \tanh \rho,
  \label{metricPoincare}
\end{align}
and the metric in terms of $\rho,y,t$ coordinates is given by
\begin{equation}
    ds^2 = d \rho^2 + \cosh^2 \rho \frac{-dt^2+dy^2}{y^2}.
    \label{metricAdS2foliation}
\end{equation}

Let us introduce the EoW brane \brane. We impose the Neumann boundary condition on \brane following \cite{Takayanagi:2011zk}:
\begin{align}
  K_{ab}=(K-T)h_{ab}. \label{metricboundarycondition}
\end{align}
We consider the following simple solution.  In the coordinates \eqref{metricAdS2foliation}, we put the EoW brane \brane at $\rho=\rho_*$, where $\rho_*$ is constant, and restrict the bulk spacetime \bulk to the region $\rho_* \le \rho$. This geometry satisfies the equations of motion and the boundary condition \eqref{metricboundarycondition} when $T$ and $\rho_*$ are related by
\begin{align}
  T=-\tanh\rho_*
\end{align}
The asymptotic boundary is located at $\rho=+\infty$.
In the calculation, we will put the cutoff at $\rho=\rho_{\infty}$ and restrict the range of $\rho$ to $\rho_*\le \rho \le \rho_{\infty}$ to regulate the divergence.  Then we will take the $\rho_{\infty}\to \infty$ limit.

\subsection{Maxwell--Chern--Simons theory}
\label{2.2}
Now we introduce the bulk action and solve its equation of motion.
We work in the probe limit and neglect the backreaction of the gauge field on the background geometry.
Its solution is used to obtain the current in the BCFT.
For an early analysis of Maxwell--Chern--Simons theory in the AdS/CFT correspondence, see \cite{Minces:1999tp}.

We consider the action of a Maxwell--Chern--Simons system \cite{Deser:1981wh,Deser:1982vy} in the bulk spacetime
\begin{equation}
\label{MCSaction}
  \action
  = - \frac{1}{4q^2} \left(
      \int_\bulk d^{3}x \sqrt{-g} \, F_{\mu\nu} F^{\mu\nu}
    -  \mass \int_\bulk d^3x \, \varepsilon^{\mu\nu\sigma} A_\mu F_{\nu\sigma} \right),
\end{equation}
where \(\varepsilon^{\mu\nu\sigma}\) is the Levi-Civita symbol and it is defined by \(\varepsilon^{\rho yt}=1\).  $F_{\mu\nu}:=\del_{\mu}A_{\nu}-\del_{\nu}A_{\mu}$ is the field strength, and $q$ is the Maxwell coupling.  $\alpha$ is a constant related to the Chern--Simons level $k$ by $\alpha:=\frac{k q^2}{2\pi}$.
The Chern--Simons term corresponds to the anomaly of the symmetry in  the BCFT.
We take \(\mass>0\) throughout this paper.

Varying the action action \eqref{MCSaction} gives the equation of motion:
\begin{equation}
\label{EoM}
  \nabla_\nu F^{\nu\mu}
+ \frac{\mass}{2}\frac{\varepsilon^{\mu\nu\lambda}}{\sqrt{-g}}F_{\nu\lambda} = 0 .
\end{equation}
We use the standard on-shell decomposition of the gauge field into a flat connection and a gauge-invariant massive vector field \cite{Gukov:2004id,Andrade:2011sx,Chang:2014jna}.
To do this, we rewrite \eqref{EoM} in terms of the Hodge dual of the field strength as
\begin{equation}
\label{EoMDF}
  \frac{\varepsilon^{\mu\nu\lambda}}{\sqrt{-g}}\,\del_\nu
  \left[
  \frac{1}{2}g_{\lambda\gamma}\frac{\varepsilon^{\gamma\sigma\tau}}{\sqrt{-g}}F_{\sigma\tau}
  \right]
  + \frac{\mass}{2}\frac{\varepsilon^{\mu\nu\lambda}}{\sqrt{-g}}F_{\nu\lambda} = 0.
\end{equation}
We define the gauge-invariant vector field \(B^\mu\) by
\begin{equation}
  B^\mu := - \frac{1}{2\mass}\frac{\varepsilon^{\mu\nu\lambda}}{\sqrt{-g}}F_{\nu\lambda}.
\end{equation}
Then \eqref{EoM} can be rewritten as a first-order differential equation
\begin{equation}
\label{eqB}
  \frac{\varepsilon^{\mu\nu\lambda}}{\sqrt{-g}}\del_\nu B_\lambda + \mass B^\mu=0.
\end{equation}
In addition, the equation \eqref{EoMDF} also implies
\begin{equation}
  F_{\mu\nu} = \del_\mu B_\nu - \del_\nu B_\mu.
\end{equation}
Using \(A^{(0)}_\mu := A_\mu-B_\mu\), the gauge field \(A_\mu\) can be expressed as \(A_\mu = A^{(0)}_\mu + B_\mu\)
and \(A^{(0)}_\mu\) is a flat connection which means \(\del_\mu A^{(0)}_\nu - \del_\nu A^{(0)}_\mu = 0\).

Employing the axial gauge \(A_\rho=0\) and the ansatz \(F_{yt}=0\), the following equations are derived from \eqref{EoM}
\begin{align}
  \partial_\rho\partial_\rho A_t - \mass \partial_\rho A_y &=0, \label{AszEoM1}\\
  \partial_\rho\partial_\rho A_y - \mass \partial_\rho A_t &=0,\\
  \partial_t \partial_\rho A_t - \partial_y \partial_\rho A_y &=0. \label{AszEoM3}
\end{align}

In this paper we impose the Dirichlet boundary condition \(A_i=0\) \((i=t,y)\) on the EoW brane.
The solutions of \eqref{AszEoM1}-\eqref{AszEoM3} are  parametrized by two arbitrary functions.
The gauge field obtained by performing a gauge transformation on the solutions is written as follows
\begin{equation}
  A_\mu = \bar{A}_\mu + \del_\mu \lambda(x), \label{AtSol}
\end{equation}
where
\begin{align}
  \bar{A}_t &= f(t-y) (e^{-\mass\rho} - e^{-\mass\rho_\ast})
             + g(t+y) (e^{\mass\rho} - e^{\mass\rho_\ast}), \notag\\
  \bar{A}_y &= - f(t-y) (e^{-\mass\rho} - e^{-\mass\rho_\ast})
               + g(t+y) (e^{\mass\rho} - e^{\mass\rho_\ast}), \notag\\
  \bar{A}_\rho &= 0. \label{AxialGaugeSol}
\end{align}
\(f(x)\) and \(g(x)\) are arbitrary functions.
\(\lambda(x)\) is a function that satisfies \(\del_i \lambda=0\) on the EoW brane to maintain the Dirichlet boundary condition.
We should consider this gauge transformation since such degrees of freedom may become physical in the presence of a Chern--Simons term on a spacetime with boundaries.

To consider the corner $\Sigma \cap Q$, we introduce the bulk magnetic current
\begin{equation}
  \label{M1current}
  j_{\mathrm m}^{\mu}:=\frac{1}{2}\varepsilon^{\mu\nu\lambda}F_{\nu\lambda},
  \qquad \del_\mu j_{\mathrm m}^{\mu}=0.
\end{equation}
Its conservation in the bulk follows from the Bianchi identity.
We require this magnetic symmetry to be preserved even in the spacetime with boundary.
In quantum gravity, it is believed that an exact symmetry must be a gauge symmetry.
We regard this magnetic symmetry as an exact gauge symmetry and thus we impose this magnetic current conservation.

Since \(F_{yt}=0\), we have \(j_{\mathrm m}^{\rho}=0\), so there is no magnetic flux through the surfaces of constant \(\rho\).
As an additional boundary condition, we require that no net bulk magnetic charge flow into or out of the corner \(y=0\), where the BCFT boundary is located:
\begin{equation}
  \lim_{\epsilon\to0^+}\int_{\rho_*}^{\rho_\infty}d\rho\,
  j_{\mathrm m}^{y}(\rho,t,\epsilon)=0.
  \label{magneticNoFlux}
\end{equation}
This condition selects a family of solutions.
Using \(j_{\mathrm m}^{y}=-\partial_\rho \bar{A}_t\) and the Dirichlet condition \(\bar{A}_t(\rho_*,t,y)=0\), we therefore obtain
\begin{equation}
  \int_{\rho_*}^{\rho_\infty}d\rho\,j_{\mathrm m}^{y}
  =-\int_{\rho_*}^{\rho_\infty}d\rho\,\partial_\rho \bar{A}_t
  =-\bar{A}_t(\rho_\infty,t,y).
\end{equation}
Thus, \eqref{magneticNoFlux} is equivalent to
\begin{equation}
  \bar{A}_t(\rho_\infty,t,0)=0.
\end{equation}
Substituting \eqref{AxialGaugeSol} into this condition gives
\begin{align}
  g(t)&=-f(t)\frac{e^{-\mass\rho_\infty}-e^{-\mass\rho_*}}
                         {e^{\mass\rho_\infty}-e^{\mass\rho_*}} \nonumber\\
      &=e^{-\mass(\rho_*+\rho_\infty)}f(t).
  \label{fgRelation}
\end{align}

Instead of the axial gauge \(\lambda(x)=0\), we consider imposing the Dirichlet boundary condition on the asymptotic boundary.
For the finite cut off we can choose the following function
\begin{equation}
  \lambda(x)= -\{F(t-y)-F(t+y)\}(e^{-\mass\rho} - e^{-\mass\rho_\ast})\chi(\rho),
\end{equation}
where \(\chi(\rho)\) is a smooth function that satisfies \(\chi(\rho)=0\) in a neighborhood of $\rho_{\ast}$ and \(\chi(\rho)=1\) in a neighborhood of $\rho_{\infty}$. \(F(x)\) is an antiderivative of \(f(x)\) so that $F'(x)=f(x)$.
Substituting \eqref{fgRelation} into \eqref{AtSol}, we obtain
\begin{align}
  A_t &= f(t-y) (e^{-\mass\rho} - e^{-\mass\rho_\ast})(1-\chi)
       + f(t+y) \{e^{\mass(\rho-\rho_\ast-\rho_\infty)} - e^{-\mass\rho_\infty} + (e^{-\mass\rho}-e^{-\mass\rho_\ast})\chi\},
  \label{AtSol2}\\
  A_y &= - f(t-y) (e^{-\mass\rho} - e^{-\mass\rho_\ast})(1-\chi)
         + f(t+y) \{e^{\mass(\rho-\rho_\ast-\rho_\infty)} - e^{-\mass\rho_\infty} + (e^{-\mass\rho}-e^{-\mass\rho_\ast})\chi\},
  \label{AySol2}\\
  A_\rho &= -\{F(t-y)-F(t+y)\}\{(e^{-\mass\rho} - e^{-\mass\rho_\ast})\chi'- \mass e^{-\mass\rho} \chi\}.
  \label{ArhoSol2}
\end{align}
These indeed satisfy \(A_i(\rho_\ast)=A_i(\rho_\infty)=0\).

\subsection{Current}
\label{2.3}
Here we obtain the current in BCFT by using the bulk gauge field \eqref{AtSol2}-\eqref{ArhoSol2},
and show that it is not conserved at the boundary of the manifold defined on BCFT with 't Hooft anomaly.
In other words, we verify that the component of the current normal to the boundary does not vanish at the boundary.

First, we derive the current in BCFT.
According to the AdS/CFT conjecture, the current dual to the bulk gauge field can be read off from the boundary term obtained by varying the renormalized action \(\Iren\) defined by adding the appropriate boundary term to the action \eqref{MCSaction}.  We employ the scheme introduced in \cite{Jensen:2010em,Andrade:2011sx}.  In this scheme, the boundary term of the renormalized variation is given by
\begin{equation}
  \delta \Iren = -\frac{\mass}{2q^2}\int_{\rho=\rho_\infty} dydt \, \{(A_t^{(0)}+A_y^{(0)})(\delta A_t^{(0)}- \delta A_y^{(0)}) + (B_t-B_y)(\delta B_t+ \delta B_y)\}.\label{deltaIren}
\end{equation}

We need the component of the current normal to the boundary in the coordinates \(t,x\), but we use the hyperbolic slicing coordinates \(\rho,y,t\) in the bulk.
So we must return to the Poincaré coordinates \(z,t,x\) to obtain the current correctly, and the \(x\) component of \(J^\mu\) is given by
\begin{align}
  \label{Jx}
  J^x = \frac{\partial x}{\partial x^\mu} J^\mu = J^y \tanh{\rho}.
\end{align}

In the variation \(\delta \Iren\), the flat connection \(A^{(0)}_i\) corresponds to the BCFT current \cite{Jensen:2010em,Andrade:2011sx}.
Then we can obtain \(J^y\) as follows
\begin{equation}
\label{Currenty}
  J^y = \frac{\delta \Iren}{\delta A^{(0)}_{y}}
      = \frac{\mass}{2\couple^2}(A_t^{(0)}+A_y^{(0)}).
\end{equation}
Before removing the cutoff, we evaluate the current at \(\rho=\rho_\infty\) by substituting the solutions \eqref{AtSol2}-\eqref{ArhoSol2} into \eqref{Currenty}:
\begin{equation}
  \left.J^y\right|_{\rho=\rho_\infty}
  =-\frac{\mass}{\couple^2}f(t+y)e^{-\mass\rho_*}.
  \label{JyCutoff}
\end{equation}
Although \(g\to0\) as \(\rho_\infty\to\infty\) at fixed \(f\) and \(\rho_*\), its contribution at the cutoff surface remains finite because \(e^{\mass\rho_\infty}g(t)=e^{-\mass\rho_*}f(t)\).
We now take \(\rho_\infty\to\infty\) at fixed \(t\) and \(x>0\).
On the cutoff surface, \eqref{metricPoincare} gives \(y=x/\tanh\rho_\infty\to x\) and \(z=x/\sinh\rho_\infty\to0\).
Using \eqref{Jx} and \(\mass>0\), we obtain the BCFT current
\begin{equation}
  J^x = -\frac{\mass}{\couple^2}f(t+x)e^{-\mass\rho_*}.
\end{equation}

Finally, after taking the asymptotic limit, we approach the BCFT boundary by sending \(x\to0^+\):
\begin{equation}
  J^x(x=0) = -\frac{\mass}{\couple^2}f(t)e^{-\mass\rho_\ast}.
\end{equation}
Indeed, we find that the current is nonzero at the boundary.

Therefore, using the AdS/BCFT correspondence, we have holographically demonstrated that the current cannot be conserved at the boundary of BCFT with ’t Hooft anomaly.

\section{Mode decomposition}
\label{3}
In Section \ref{2}, we computed the boundary current using an explicit family of solutions satisfying \(F_{yt}=0\).
We now relax this restriction to describe the bulk modes and identify how that family fits into the broader solution space.

The general solution of \eqref{EoM} is obtained by solving \eqref{eqB} for \(B_\mu\) and adding a flat connection \(A^{(0)}_\mu\).
Acting with the curl operator once more on \eqref{eqB} shows that \(B_\mu\) satisfies the Proca equation.
The general structure of solutions to the Proca equation in this AdS/BCFT geometry was studied in \cite{Suzuki:2024cqy}.

We now solve \eqref{eqB} explicitly.
For convenience, we write \(a(\rho):=\cosh\rho\) for the warp factor in \eqref{metricAdS2foliation}.
Since we have time translation symmetry in this background, it is enough to consider \(B_\mu\) as in \(B_\mu=e^{-i \omega t}b_\mu(\rho,y)\). Then, we can reduce \eqref{eqB} to differential equations in two variables
\begin{align}
  \mass b_y - i \omega b_\rho - \del_\rho b_t &=0, \label{FOEoM1}\\
  \mass b_t - \del_\rho b_y + \del_y b_\rho &=0, \label{FOEoM2}\\
  \mass b_\rho + \frac{y^2}{a^2} (\del_y b_t + i \omega b_y) &=0. \label{FOEoM3}
\end{align}
Then by solving \eqref{FOEoM1}-\eqref{FOEoM3} for \(b_\mu\), we can determine \(B_\mu\).

For \(b_\rho\), we can derive a second-order differential equation from \eqref{FOEoM1}-\eqref{FOEoM3}
\begin{equation}
\label{b_rho}
  (- \del_\rho^2 + \mass^2)(a^2 b_\rho) = y^2 (\del_y^2 + \omega^2) b_\rho.
\end{equation}
If we assume the separable form \(b_\rho=a^{-1}\Rcal(\rho)Y(y)\), \eqref{b_rho} separates into the two equations:
\begin{align}
  \del_\rho^2 \Rcal + 2\tanh{\rho}\,\del_\rho \Rcal + (1-\mass^2)\Rcal &= -\frac{\lambda}{a^2}\Rcal, \label{Req}\\
  \del_y^2 Y + \omega^2 Y - \frac{\lambda}{y^2}Y &= 0. \label{Yeq}
\end{align}
Here the factor of \(a^{-1}\) in \(b_\rho\) is factored out to match the convention used in \cite{Suzuki:2024cqy}.
These can be solved by
\begin{align}
  \Rcal(\rho) &= (1-\zeta^2)^{\frac{1}{2}}\left[
    c_1 P_{\nu-\frac{1}{2}}^\mu(\zeta) + c_2 Q_{\nu-\frac{1}{2}}^\mu(\zeta)
  \right],\\
  Y(y) &= y^{\frac{1}{2}} \left[
    c_3 J_\nu(|\omega|y) + c_4 Y_\nu(|\omega|y)
  \right],
\end{align}
where
\begin{equation}
  \zeta=\tanh\rho, \quad \mu=|\mass|, \quad \nu=\frac{1}{2}\sqrt{1+4\lambda}\,.
\end{equation}
\(P\) and \(Q\) denote the associated Legendre functions, and \(J_\nu\) and \(Y_\nu\) denote Bessel functions of the first and second kinds, respectively.
For further details, see \cite{Suzuki:2024cqy}.

Finally, we consider the remaining \(b_y\) and \(b_t\).
From \eqref{FOEoM1} and \eqref{FOEoM2} we obtain the following equations
\begin{align}
  \del_\rho (b_y - b_t) + \mass (b_y - b_t) &= \del_y b_\rho + i \omega b_\rho, \\
  \del_\rho (b_y + b_t) - \mass (b_y + b_t) &= \del_y b_\rho - i \omega b_\rho.
\end{align}
Then the general solutions for \(b_y-b_t\) and \(b_y+b_t\) are
\begin{align}
  b_y-b_t &= \left(C_1(y)+(\del_y + i \omega)Y
    \int_{\rho_1}^\rho d\rho' e^{\mass\rho'}a(\rho')^{-1}\Rcal(\rho')
    \right)e^{-\mass\rho}, \label{y-t}\\
  b_y+b_t &= \left(C_2(y)+(\del_y - i\omega)Y
    \int_{\rho_2}^\rho d\rho'e^{-\mass\rho'}a(\rho')^{-1}\Rcal(\rho')
    \right)e^{\mass\rho}. \label{y+t}
\end{align}
For \(\lambda\neq0\), the integrals can be evaluated using the following identities:
\begin{align}
  \pdv{\rho} \left\{\frac{1}{\lambda}e^{\mass\rho}(\mass-\del_\rho)(a\Rcal)\right\} 
  &= e^{\mass\rho}a^{-1}\Rcal
  = \pdv{\rho} \left\{ \int_{\rho_1}^\rho d\rho' e^{\mass\rho'}a(\rho')^{-1}\Rcal(\rho') \right\}, \\
  \pdv{\rho} \left\{-\frac{1}{\lambda}e^{-\mass\rho}(\mass+\del_\rho)(a\Rcal)\right\}
  &= e^{-\mass\rho}a^{-1}\Rcal
  = \pdv{\rho}\left\{\int_{\rho_2}^\rho d\rho'e^{-\mass\rho'}a(\rho')^{-1}\Rcal(\rho')\right\}.
\end{align}
Evaluating the integrals in \eqref{y-t} and \eqref{y+t} and imposing the remaining equation \eqref{FOEoM3}, we obtain, for \(\lambda\neq0\),
\begin{align}
  b_y&= C_3(\omega)e^{-\mass\rho}e^{i\omega y}+C_4(\omega)e^{\mass\rho}e^{-i\omega y}
    +\frac{1}{\lambda}\left(
    i\omega\mass(a\Rcal)Y-(\del_\rho(a\Rcal))\del_y Y\right), \\
  b_t&= -C_3(\omega)e^{-\mass\rho}e^{i\omega y}+C_4(\omega)e^{\mass\rho}e^{-i\omega y}
    +\frac{1}{\lambda}\left(
    i\omega(\del_\rho(a\Rcal))Y-\mass(a\Rcal)\del_y Y\right),
\end{align}
where \(C_3(\omega)\) and \(C_4(\omega)\) are arbitrary constans.
The \(\lambda=0\) solutions can be recovered by taking \(\lambda\to0\) along smooth families of \(\Rcal\) and \(Y\).
Any \(1/\lambda\) poles can be canceled by suitable \(\lambda\)-dependent choices of the homogeneous coefficients \(c_1,\dots,c_4\).


Upon inverse Fourier transformation, the homogeneous factors \(e^{-i\omega t}e^{\pm i\omega y}\) give arbitrary functions of \(t\mp y\).
Absorbing the overall sign of the \(C_3(\omega)\) branch into \(f\), we denote these functions by \(f(t-y)\) and \(g(t+y)\).
Superposing the reconstructed modes and adding the flat connection then gives
\begin{align}
  A_\rho &= A_\rho^{(0)}+ \int d\omega\, e^{-i\omega t} \sum_\lambda a^{-1}\Rcal Y , \label{GArho}\\
  A_y &= A_y^{(0)} - f(t-y) e^{-\mass\rho} + g(t+y) e^{\mass\rho} - (\mass\del_t+\del_\rho\del_y) \int d\omega e^{-i\omega t} \sum_\lambda \frac{1}{\lambda} a \Rcal Y, \label{GAy}\\
  A_t &= A_t^{(0)} + f(t-y) e^{-\mass\rho} + g(t+y) e^{\mass\rho} - (\del_t \del_\rho+\mass\del_y) \int d\omega e^{-i\omega t} \sum_\lambda \frac{1}{\lambda} a \Rcal Y. \label{GAt}
\end{align}
The allowed mode coefficients and the spectrum of \(\lambda\) are determined by boundary conditions; these have not yet been imposed on the additional modes.

Let us now recover the solutions used to compute the boundary current in Section \ref{2}.
The definition of \(B_\mu\) implies that \(F_{yt}=0\) is equivalent to \(B_\rho=0\).
We therefore set the modes reconstructed from \(b_\rho\) to zero in \eqref{GArho}--\eqref{GAt}, retaining the terms proportional to \(f(t-y)\) and \(g(t+y)\).
Choosing the flat connection to be
\begin{align}
  A_\rho^{(0)} &= \del_{\rho}\lambda, \nonumber\\
  A_t^{(0)} &= -f(t-y)e^{-\mass\rho_*}-g(t+y)e^{\mass\rho_*}+\del_t \lambda, \nonumber\\
  A_y^{(0)} &= f(t-y)e^{-\mass\rho_*}-g(t+y)e^{\mass\rho_*}+\del_y \lambda,
  \label{flatConnectionExplicit}
\end{align}
then reproduces the solution \eqref{AtSol} with \eqref{AxialGaugeSol}.
Indeed, \eqref{flatConnectionExplicit} satisfies \(\partial_t A_y^{(0)}-\partial_y A_t^{(0)}=0\), and its tangential components cancel those of \(B_\mu\) at \(\rho=\rho_*\), ensuring the Dirichlet condition on the EoW brane.
The relation between \(f\) and \(g\) follows upon imposing the additional magnetic-flux condition of Section \ref{2.2}.


\subsection*{Acknowledgement}
The work of S.Y. was supported in part by JSPS KAKENHI Grant Number 26K07086.
The work of H.T. was supported in part by JST SPRING, Grant Number JPMJSP2138.
\bibliographystyle{utphys}
\bibliography{ref}

\providecommand{\href}[2]{#2}\begingroup\raggedright\begin{thebibliography}{10}

\bibitem{Maldacena:1997re}
J.~M. Maldacena, ``{The Large $N$ limit of superconformal field theories and supergravity},'' \href{https://dx.doi.org/10.4310/ATMP.1998.v2.n2.a1}{{\em Adv. Theor. Math. Phys.} {\bfseries 2} (1998) 231--252}, \href{https://arxiv.org/abs/hep-th/9711200}{{\ttfamily arXiv:hep-th/9711200}}.

\bibitem{Cardy:1984bb}
J.~L. Cardy, ``{Conformal Invariance and Surface Critical Behavior},'' \href{https://dx.doi.org/10.1016/0550-3213(84)90241-4}{{\em Nucl. Phys. B} {\bfseries 240} (1984) 514--532}.

\bibitem{Cardy:1989ir}
J.~L. Cardy, ``{Boundary Conditions, Fusion Rules and the Verlinde Formula},'' \href{https://dx.doi.org/10.1016/0550-3213(89)90521-X}{{\em Nucl. Phys. B} {\bfseries 324} (1989) 581--596}.

\bibitem{Cardy:2004hm}
J.~L. Cardy, ``{Boundary conformal field theory},'' \href{https://arxiv.org/abs/hep-th/0411189}{{\ttfamily arXiv:hep-th/0411189}}.

\bibitem{McAvity:1993ue}
D.~M. McAvity and H.~Osborn, ``{Energy momentum tensor in conformal field theories near a boundary},'' \href{https://dx.doi.org/10.1016/0550-3213(93)90005-A}{{\em Nucl. Phys. B} {\bfseries 406} (1993) 655--680}, \href{https://arxiv.org/abs/hep-th/9302068}{{\ttfamily arXiv:hep-th/9302068}}.

\bibitem{McAvity:1995zd}
D.~M. McAvity and H.~Osborn, ``{Conformal field theories near a boundary in general dimensions},'' \href{https://dx.doi.org/10.1016/0550-3213(95)00476-9}{{\em Nucl. Phys. B} {\bfseries 455} (1995) 522--576}, \href{https://arxiv.org/abs/cond-mat/9505127}{{\ttfamily arXiv:cond-mat/9505127}}.

\bibitem{Takayanagi:2011zk}
T.~Takayanagi, ``{Holographic Dual of BCFT},'' \href{https://dx.doi.org/10.1103/PhysRevLett.107.101602}{{\em Phys. Rev. Lett.} {\bfseries 107} (2011) 101602}, \href{https://arxiv.org/abs/1105.5165}{{\ttfamily arXiv:1105.5165 [hep-th]}}.

\bibitem{Fujita:2011fp}
M.~Fujita, T.~Takayanagi, and E.~Tonni, ``{Aspects of AdS/BCFT},'' \href{https://dx.doi.org/10.1007/JHEP11(2011)043}{{\em JHEP} {\bfseries 11} (2011) 043}, \href{https://arxiv.org/abs/1108.5152}{{\ttfamily arXiv:1108.5152 [hep-th]}}.

\bibitem{Jensen:2017eof}
K.~Jensen, E.~Shaverin, and A.~Yarom, ``{{\textquoteright}t Hooft anomalies and boundaries},'' \href{https://dx.doi.org/10.1007/JHEP01(2018)085}{{\em JHEP} {\bfseries 01} (2018) 085}, \href{https://arxiv.org/abs/1710.07299}{{\ttfamily arXiv:1710.07299 [hep-th]}}.

\bibitem{Thorngren:2020yht}
R.~Thorngren and Y.~Wang, ``{Anomalous symmetries end at the boundary},'' \href{https://dx.doi.org/10.1007/JHEP09(2021)017}{{\em JHEP} {\bfseries 09} (2021) 017}, \href{https://arxiv.org/abs/2012.15861}{{\ttfamily arXiv:2012.15861 [hep-th]}}.

\bibitem{Choi:2023xjw}
Y.~Choi, B.~C. Rayhaun, Y.~Sanghavi, and S.-H. Shao, ``{Remarks on boundaries, anomalies, and noninvertible symmetries},'' \href{https://dx.doi.org/10.1103/PhysRevD.108.125005}{{\em Phys. Rev. D} {\bfseries 108} no.~12, (2023) 125005}, \href{https://arxiv.org/abs/2305.09713}{{\ttfamily arXiv:2305.09713 [hep-th]}}.

\bibitem{Witten:1998qj}
E.~Witten, ``{Anti de Sitter space and holography},'' \href{https://dx.doi.org/10.4310/ATMP.1998.v2.n2.a2}{{\em Adv. Theor. Math. Phys.} {\bfseries 2} (1998) 253--291}, \href{https://arxiv.org/abs/hep-th/9802150}{{\ttfamily arXiv:hep-th/9802150}}.

\bibitem{Jensen:2010em}
K.~Jensen, ``{Chiral anomalies and AdS/CMT in two dimensions},'' \href{https://dx.doi.org/10.1007/JHEP01(2011)109}{{\em JHEP} {\bfseries 01} (2011) 109}, \href{https://arxiv.org/abs/1012.4831}{{\ttfamily arXiv:1012.4831 [hep-th]}}.

\bibitem{Callan:1984sa}
C.~G. Callan, Jr. and J.~A. Harvey, ``{Anomalies and Fermion Zero Modes on Strings and Domain Walls},'' \href{https://dx.doi.org/10.1016/0550-3213(85)90489-4}{{\em Nucl. Phys. B} {\bfseries 250} (1985) 427--436}.

\bibitem{Suzuki:2024cqy}
K.~Suzuki, ``{Gauge symmetries and conserved currents in AdS/BCFT},'' \href{https://dx.doi.org/10.1007/JHEP06(2024)137}{{\em JHEP} {\bfseries 06} (2024) 137}, \href{https://arxiv.org/abs/2403.07325}{{\ttfamily arXiv:2403.07325 [hep-th]}}.

\bibitem{Fujita:2012fp}
M.~Fujita, M.~Kaminski, and A.~Karch, ``{SL(2,Z) Action on AdS/BCFT and Hall Conductivities},'' \href{https://dx.doi.org/10.1007/JHEP07(2012)150}{{\em JHEP} {\bfseries 07} (2012) 150}, \href{https://arxiv.org/abs/1204.0012}{{\ttfamily arXiv:1204.0012 [hep-th]}}.

\bibitem{Miao:2022oas}
R.-X. Miao and Y.-Q. Zeng, ``{Holographic anomalous chiral current near a boundary},'' \href{https://dx.doi.org/10.1016/j.physletb.2023.137700}{{\em Phys. Lett. B} {\bfseries 838} (2023) 137700}, \href{https://arxiv.org/abs/2210.05203}{{\ttfamily arXiv:2210.05203 [hep-th]}}.

\bibitem{Zhao:2023rno}
T.-M. Zhao and R.-X. Miao, ``{General holographic chiral theory near a boundary},'' \href{https://dx.doi.org/10.1016/j.physletb.2023.138383}{{\em Phys. Lett. B} {\bfseries 848} (2024) 138383}.

\bibitem{Gibbons:1976ue}
G.~W. Gibbons and S.~W. Hawking, ``{Action Integrals and Partition Functions in Quantum Gravity},'' \href{https://dx.doi.org/10.1103/PhysRevD.15.2752}{{\em Phys. Rev. D} {\bfseries 15} (1977) 2752--2756}.

\bibitem{York:1972sj}
J.~W. York, Jr., ``{Role of conformal three geometry in the dynamics of gravitation},'' \href{https://dx.doi.org/10.1103/PhysRevLett.28.1082}{{\em Phys. Rev. Lett.} {\bfseries 28} (1972) 1082--1085}.

\bibitem{Skenderis:2002wp}
K.~Skenderis, ``{Lecture notes on holographic renormalization},'' \href{https://dx.doi.org/10.1088/0264-9381/19/22/306}{{\em Class. Quant. Grav.} {\bfseries 19} (2002) 5849--5876}, \href{https://arxiv.org/abs/hep-th/0209067}{{\ttfamily arXiv:hep-th/0209067}}.

\bibitem{Minces:1999tp}
P.~Minces and V.~O. Rivelles, ``{Chern-Simons theories in the AdS / CFT correspondence},'' \href{https://dx.doi.org/10.1016/S0370-2693(99)00411-6}{{\em Phys. Lett. B} {\bfseries 455} (1999) 147--154}, \href{https://arxiv.org/abs/hep-th/9902123}{{\ttfamily arXiv:hep-th/9902123}}.

\bibitem{Deser:1981wh}
S.~Deser, R.~Jackiw, and S.~Templeton, ``{Topologically Massive Gauge Theories},'' \href{https://dx.doi.org/10.1016/0003-4916(82)90164-6}{{\em Annals Phys.} {\bfseries 140} (1982) 372--411}. [Erratum: Annals Phys. 185, 406 (1988)].

\bibitem{Deser:1982vy}
S.~Deser, R.~Jackiw, and S.~Templeton, ``{Three-Dimensional Massive Gauge Theories},'' \href{https://dx.doi.org/10.1103/PhysRevLett.48.975}{{\em Phys. Rev. Lett.} {\bfseries 48} (1982) 975--978}.

\bibitem{Gukov:2004id}
S.~Gukov, E.~Martinec, G.~W. Moore, and A.~Strominger, \href{https://dx.doi.org/10.1142/9789812775344_0036}{``{Chern-Simons gauge theory and the AdS(3) / CFT(2) correspondence},''} in {\em {From Fields to Strings: Circumnavigating Theoretical Physics: A Conference in Tribute to Ian Kogan}}, pp.~1606--1647.
\newblock 3, 2004.
\newblock \href{https://arxiv.org/abs/hep-th/0403225}{{\ttfamily arXiv:hep-th/0403225}}.

\bibitem{Andrade:2011sx}
T.~Andrade, J.~I. Jottar, and R.~G. Leigh, ``{Boundary Conditions and Unitarity: the Maxwell-Chern-Simons System in $AdS_3/CFT_2$},'' \href{https://dx.doi.org/10.1007/JHEP05(2012)071}{{\em JHEP} {\bfseries 05} (2012) 071}, \href{https://arxiv.org/abs/1111.5054}{{\ttfamily arXiv:1111.5054 [hep-th]}}.

\bibitem{Chang:2014jna}
H.-C. Chang, M.~Fujita, and M.~Kaminski, ``{From Maxwell-Chern-Simons theory in $AdS_3$ towards hydrodynamics in $1+1$ dimensions},'' \href{https://dx.doi.org/10.1007/JHEP10(2014)118}{{\em JHEP} {\bfseries 10} (2014) 118}, \href{https://arxiv.org/abs/1403.5263}{{\ttfamily arXiv:1403.5263 [hep-th]}}.

\end{thebibliography}\endgroup
\end{document}